\documentclass[a4paper,fleqn,usenatbib]{mnras}

\usepackage[T1]{fontenc}
\usepackage{ae,aecompl}

\usepackage{graphicx}	
\usepackage{amsmath}	
\usepackage{amssymb}	

\title[The Milky Way globular cluster system]{Characteristic radii of the Milky Way Globular Clusters
}

\author[A.E. Piatti et al.]{
Andr\'es E. Piatti$^{1,2}$\thanks{E-mail: andres.piatti@unc.edu.ar}, Jeremy Webb$^{3}$ and
Raymond Carlberg$^3$\\
$^{1}$Consejo Nacional de Investigaciones Cient\'{\i}ficas y T\'ecnicas, Godoy Cruz 2290, C1425FQB, 
Buenos Aires, Argentina\\
$^{2}$Observatorio Astron\'omico de C\'ordoba, Laprida 854, 5000, 
C\'ordoba, Argentina\\
$^{3}$Department of Astronomy and Astrophysics, University of Toronto, 50 St. George Street, Toronto, ON, M5S 3H4, Canada
}

\date{Accepted XXX. Received YYY; in original form ZZZ}

\pubyear{2019}

\begin{document}
\label{firstpage}
\pagerange{\pageref{firstpage}--\pageref{lastpage}}
\maketitle

\begin{abstract}
We report on the extent of the effects of the Milky Way's gravitational field in
shaping the structural parameters and internal dynamics
of its globular cluster population. We make use of a homogeneous,
up-to-date data set with kinematics, structural properties,
current and initial masses of 156 globular clusters. In general, cluster radii increase
as the Milky Way potential weakens; with the core and Jacobi radii being those 
which increase at the slowest and fastest rate respectively. We
interpret this result as the innermost regions of globular clusters being less sensitive to changes in the tidal forces with the Galactocentric distance. The Milky Way's gravitational field also seems to have differentially accelerated the internal dynamical evolution of individual clusters, with those toward the bulge appearing dynamically older. Finally we
find a sub-population consisting of both compact and extended globular clusters (as defined by their $r_h/r_J$ ratio) beyond 8 kpc that appear to have lost a large fraction of their initial mass lost via disruption. Moreover, we identify a third group with $r_h/r_J$ > 0.4, which have lost an even larger fraction of their initial mass by disruption. In both cases the high fraction of mass lost is likely due to their large orbital eccentricities and inclination angles, which lead to them experiencing more tidal shocks at perigalacticon and during disc crossings. Comparing the structural and orbital parameters of individual clusters allows for constraints to be placed on whether or not their evolution was relaxation or tidally dominated.

\end{abstract} 




\section{Introduction}

It is widely accepted that Milky Way globular clusters have lost most of their masses
through three main processes, namely: stellar evolution, two-body relaxation and tidal heating caused by the Milky Way's
gravitational field \citep{gnedinetal1999,fz01,gielesetal2008,webbetal2013,webbetal2014,brockampetal2014,alessandrinietal2014}. Stellar evolution is most important during the first few
hundred million years, while two-body relaxation becomes important as the mass loss rate due to stellar evolution continues to decrease \citep{henon1961,gielesetal2011,hh03,shukirgaliyevetal2018}. Whether or not a cluster is strongly affected by the tidal field depends on its size, mass, and orbit within the Galaxy. Since the strength of the Milky Way's tidal field weakens with galactocentric distance, it is expected that mass loss due to tidal effects follows a similar trend. Hence, the amount of mass lost by
very distant globular clusters should be less than those moving in the Milky Way bulge. Furthermore for a given orbit, clusters that are more massive or more compact will have a stronger self-gravity and will be less affected by tidal heating.
Such behaviour has been observed within the old globular clusters of the
Large Magellanic Cloud, where the galaxy's gravitational potential seems to act
differently as a function of the cluster distance from the galaxy's centre \citep{pm2018}.

From a purely theoretical approach, the literature is rich with studies on how star clusters are affected by the tidal field of their host galaxy 
\citep[e.g.][]{gnedinetal1999,lamersetal2005a,hh03,renaudetal2011,bm2003,gielesetal2008,km2009,webbetal2013,webbetal2014}. Most applicable to this study is the work done by \citet{bm2003}, who performed N-body simulations to study how a spherically symmetric external tidal field, in addition to stellar evolution and two-body relaxation, affects the evolution of a star cluster's mass function. They found that mass segregation leads to low-mass stars being preferentially stripped from star clusters, which causes mass functions that initial increase towards the low-mass end to begin decreasing towards the low-mass end 
\citep[see][for a direct comparison to observations]{webbetal2017}. From their suite 
of simulations, \citet{bm2003} were also able to generate a relation for estimating a cluster's dissolution time based on its orbit in the galaxy. Expectedly, inner region clusters are expected to dissolve faster than outer region clusters. A secondary dependence exists in the form of the cluster's orbital eccentricity, with clusters that have high orbital eccentricities reaching dissolution faster than clusters with low orbital eccentricities but comparable apocenters. It should be noted that the simulations were limited to 
approximately $10^5$ particles. More massive clusters will have less two body relaxation, $\propto \log{N}/N$. Tidal heating, which depends on the cluster size, has no dependence on particle mass.

For cluster's in non-spherically symmetric tidal fields, tidal heating can also occur in the form of shocks. \citet{spitzer1958} was the first to study the disruption of clusters due to tidal shocks, and found the
amount of mass lost mainly depends on the strength of the shock and the cluster`s density within its half-mass radius. Giant molecular clouds (GMCs) are now known to be the dominant source of mass loss over a cluster's lifetime 
\citep{gielesetal2006,lamersetal2006,kruijssenetal2011,gr2016}, primarily affecting clusters when they first form and the local GMC encounter rate is high. Focussing on disc shocks in particular, studies have shown that repeated passages through the Galactic disc will accelerate a cluster's dissolution time relative to a cluster orbiting in the plane of the disc \citep{go1997,gielesetal2007,donghia2010, kruijssenetal2011, webbetal2014}. Other forms of substructure, including galaxy merger-induced structure 
\citep{kruijseenetal2012} and dark matter sub-halos \citep{webbetal2019} can also subject a cluster to a tidal shock.

Finally, an additional factor that makes it difficult to estimate a cluster's mass-loss history based on its orbit alone is the fact that a significant fraction of the Galactic cluster population has been accreted \citep{massarietal2019}. Hence these clusters have not been on their current orbit for their entire lifetime, and could have lost significantly more (or less) mass than one would predict given their current orbit and structural properties. Luckily the structural properties of accreted clusters are expected to respond to the tidal field of their new host, in this case the Milky Way, very quickly \citep{miholicsetal2014}. 

Historically, studies of how Galactic globular clusters have been affected by the Milky Way have been limited by the fact that their orbits were unknown. Hence conclusions could only be drawn based on their observed properties and their present galactocentric distance. Furthermore, estimates of a cluster's mass and size were taken from a wide range of inhomogenous studies. As the proper motions of select globular clusters started becoming available, a proper analysis of how individual Galactic globular clusters have been affected by tides was possible \citep{dinescuetal1999,km2009,webbetal2012,bg2018}. In some cases, it was even possible to determine if a cluster's evolution was dominated by two-body relaxation or tidal stipping \citep{dinescuetal1999}. Unfortunately, similar to catalogues of globular cluster sturctural parameters, estimates of cluster orbits were inhomogeneous and typically incomplete. Hence it has been difficult for past studies to identify how strongly a cluster's evolution has been affected by the tidal field of its host galaxy.

With the most recent data release from the European Space Agency's {\it Gaia} satellite (DR2), the orbit of every Galactic globular cluster is now known
\citep{gaiaetal2018a,vasiliev2019,baumgardtetal2019}. Combining data from {\it Gaia} DR2 with ground-based line-of-sight velocities and estimates of each cluster's mass and size 
\citep{bh2018}, \citet{baumgardtetal2019} was able to derive the mass lost by each globular cluster since formation by integrating their orbits backwards in time and accounting for dynamical friction. 

The \citet{baumgardtetal2019} catalogue, which contains cluster positions, space velocities, orbital motion 
parameters, structural properties, and initial masses of almost all known Milky Way globular clusters  
\citep[][2010 December edition]{harris1996} is the most suitable database  for studying the extent that 
the observed structural properties of clusters have been affected by the Milky Way's tidal field. 
Hence the purpose of this study is to make use of the \citet{baumgardtetal2019}  catalogue to 
analyse how the Milky Way's gravitational field affects various observable properties of globular clusters. The estimated fraction of cluster mass lost by disruption serves as a tracer of how strongly a cluster is tidally affected, which can then be compared to several structural and dynamical parameters in order to help constrain a cluster's origin ({\it in-situ} vs accreted), its properties at formation, and whether or not its evolution is dominated by internal or external processes.

The catalogue allows for a homogeneous comparison of a cluster's structural and orbital properties, while also providing an estimate of what each cluster's mass was at formation. This type of study is only possible because Galactic globular clusters are born at the earliest epochs of the Milky Way formation \citep{Kruijssen2014, kruijssenetal2018} and, consequently, their mass lost 
due to evolutionary effects is in practice the same \citep[e.g.][]{lamersetal2005a}.
Therefore, it can be expected that any difference between their structural parameters
could be due to the difference tidal forces acting on them.

In Section 2 we introduces the data set used in this work. In Section 3, we explore how the estimated fraction of mass lost by each cluster is related to its orbit, several tracers of cluster size, its density profile, dynamical age, and tidal filling factor. Finally, in Section 4 we summarise the main conclusions of this work.

\section{Data handling}

To explore how strongly tides have shaped a cluster's evolution, we make use of the Galactic positions ($X,Y,Z$), space velocities ($U,V,W$), perigalactic
($R_{peri}$) and apogalactic ($R_{apo}$) distances, and initial ($M_{ini}$) and current 
($M_{GC}$) masses of 156 Milky Way globular clusters as derived by \citet{baumgardtetal2019}. In their study, \citet{baumgardtetal2019} estimates cluster positions and velocities using data from {\it Gaia} DR2. $R_{peri}$ and $R_{apo}$ are then determined by integrating the cluster's orbits assuming the 
\citet{irrgangetal2013} model of the Milky Way. The structural properties of each cluster, mainly their core ($r_c$) and half-mass ($r_h$) radius, are taken from \citet{bh2018} who estimate the values by comparing the density profiles of Galactic globular clusters to a large suite of direct $N$-body star cluster simulations. Since the set of orbital and structural cluster properties have been derived
by applying the same methodology, the catalogue represents the largest homogeneous, up-to-date data set of the Milky Way globular cluster system. It is useful to note that \citet{baumgardtetal2019} uses \citet{irrgangetal2013}'s model for the Milky Way in order to solve for each cluster's $R_{peri}$ and  $R_{apo}$. This model consists of a Plummer sphere bulge, a \citet{mn1975} disc, and a modified \citet{as1991} dark matter halo. We also performed an independent orbit integration for each cluster using the commonly cited MWPotential2014 Galactic model \citep{Bovy2015}, which onsists of a bulge that is represented by a spherical power-law potential, a \citet{mn1975} disc, and a \citet{navarro1997} halo. A comparison revelaed only minor differences between the orbital parameters of clusters orbiting in the innermost regions of the Milky Way (which is poorly constrained).

It is important to note that the $M_{ini}$ values in \citet{baumgardtetal2019} were obtained by integrating each cluster's 
orbit backwards in time from their observed positions and space velocities and measured
currrent  masses, taking into consideration the dynamical drag force. Using the formalism of \citet{bm2003}, the authors were able to estimate the mass loss via tidal stripping. It was additinoally assumed that clusters lose half of their
$M_{ini}$ due to stellar evolution during their first gigayear \citep[see also][]{degrijetal2005}. \citet{baumgardtetal2019} iterated over a wide range of $M_{ini}$ values  until they were able to recover each cluster's $M_{GC}$, on the basis of a linear mass loss dependence with time in a spherically symmetric, isothermal galaxy potential over the entire
age of each cluster \citep[see also][]{lamersetal2010}. One should allow for a dispersion in
the fraction of mass lost by disruption of $\sim$ 15-20 per cent due to the uncertainty in the cluster's ages \citep{helmietal2018, pfefferetal2018}. $M_{ini}$ should also be taken to be a lower limit for several reasons. First, it is does not take into account early mass lost by the cluster before it leaves its formation environment and reaches its present day orbit \citep{Kruijssen2015}. Second, interactions with GMCs, which have been shown to be the dominant source of cluster mass loss at early times \citep{gielesetal2006,kruijssenetal2011}, are also not included in the calculation). Note that
accreted clusters were likely not harassed as much by GMCs  as {\it in-situ} clusters, while
clusters formed in the inner part  of the Milky Way likely encountered more GMCs than those formed in
the outskirts. Notice also that the strength of the tidal field is expected to increase, on average, with cosmic time as the Milky Way grows \citep{renaudetal2017}, and that the presence of the bar \citep{rossietal2018} and spiral arms \citep{gielesetal2007b} will accelerate mass loss as well, which is not considered in  \citet{baumgardtetal2019}. Therefore, the estimates of initial mass used in this work are lower limits as many mass-loss 
mechanisms are not included. In fact, initial mass estimates are perhaps more accurately tracers of the tidal field strength associated with a cluster's current orbit in the present day potential of the Milky Way.

From the above data set, we computed the semi-major axis of each globular cluster`'s orbit:

\begin{equation}
a =   \frac{R_{peri} +  R_{apo}}{2}.
\end{equation}

The semi-major axis has the advantage of having no time-dependence, as opposed to the cluster`s current Galactocentric distance, and is more representative of the mean orbital distance of the globular clusters than $R_{peri}$ or $R_{apo}$.
We also computed the orbital eccentricity ($\epsilon$) as:

\begin{equation}
\epsilon = \frac{R_{apo} - R_{peri}}{R_{apo} + R_{peri}},
\end{equation}

\noindent and the values of Jacobi radii from the 
expression \citep{cw90}:

\begin{equation}
r_J = \left( \frac{M_{GC}}{3 M_{MW}} \right)^{1/3}\times a,
\end{equation}

\noindent where $M_{MW}$ is the Milky Way mass contained within the semi-major axis of the cluster's orbit, which is different for each cluster. In order to estimate $M_{MW}$ we used the same 
\citet{irrgangetal2013}
Milky Way mass profile that \citet{baumgardtetal2019} used to solve the orbit of each cluster.
We note that $a$ and $e$ are calculated in a galaxy model that is fitted to the present day properties 
of the Milky Way. However, the Milky Way has been grown over time, so that $a$ and $e$ have not
been the same over the course of a cluster's lifetime.

Finally, to estimate how much clusters have been disrupted due to relaxation and tidal heating, we split the difference between $M_{ini}$ and $M_{GC}$ up between mass lost via stellar evolution ($M_{ev}$) and mass lost due to disruption ($M_{dis}$):
 
 \begin{equation}
 M_{ini} = M_{GC} + M_{ev} + M_{dis}, 
 \end{equation}

 \noindent with $M_{ev}$ = 0.5$\times$$M_{ini}$, from which we get:
 
\begin{equation}
M_{dis}/M_{ini} = 1/2  - M_{GC}/M_{ini} .
\end{equation}

\noindent Hence eq.(5) explicitly gives the fraction of cluster mass lost by relaxation and tidal heating as computed (but not tabulated) by \citet{baumgardtetal2019}. In their study, the authors'  only considered the relationship between $M_{GC}$,
$ M_{ini}$ and Galactocentric distance (see their Figure 7). In the subsequent analysis we
use $M_{dis}/M_{ini}$ as   the indicator for tidal field strength.

We estimated the uncertainties of each derived parameter $f(x_1,x_2,...,x_n)$
on the basis of Monte Carlo simulations. We run one thousand calculations of 
$f(x_1,x_2,...,x_n)$ for each globular cluster with random distributions of each
independent variable $x_i$ over the interval $x_i \pm \sigma(x_i)$, for $i=1,...,n$, where
$\sigma(x_i)$ is the error associated to $x_i$ according to \citet{baumgardtetal2019}.
From the resulting distribution of the one thousand generated values, we
set the uncertainty of each parameter equal to the range encompassing the central 16$\%$ and 84$\%$ points.

\section{Analysis and discussion}

\begin{figure}
     \includegraphics[width=\columnwidth]{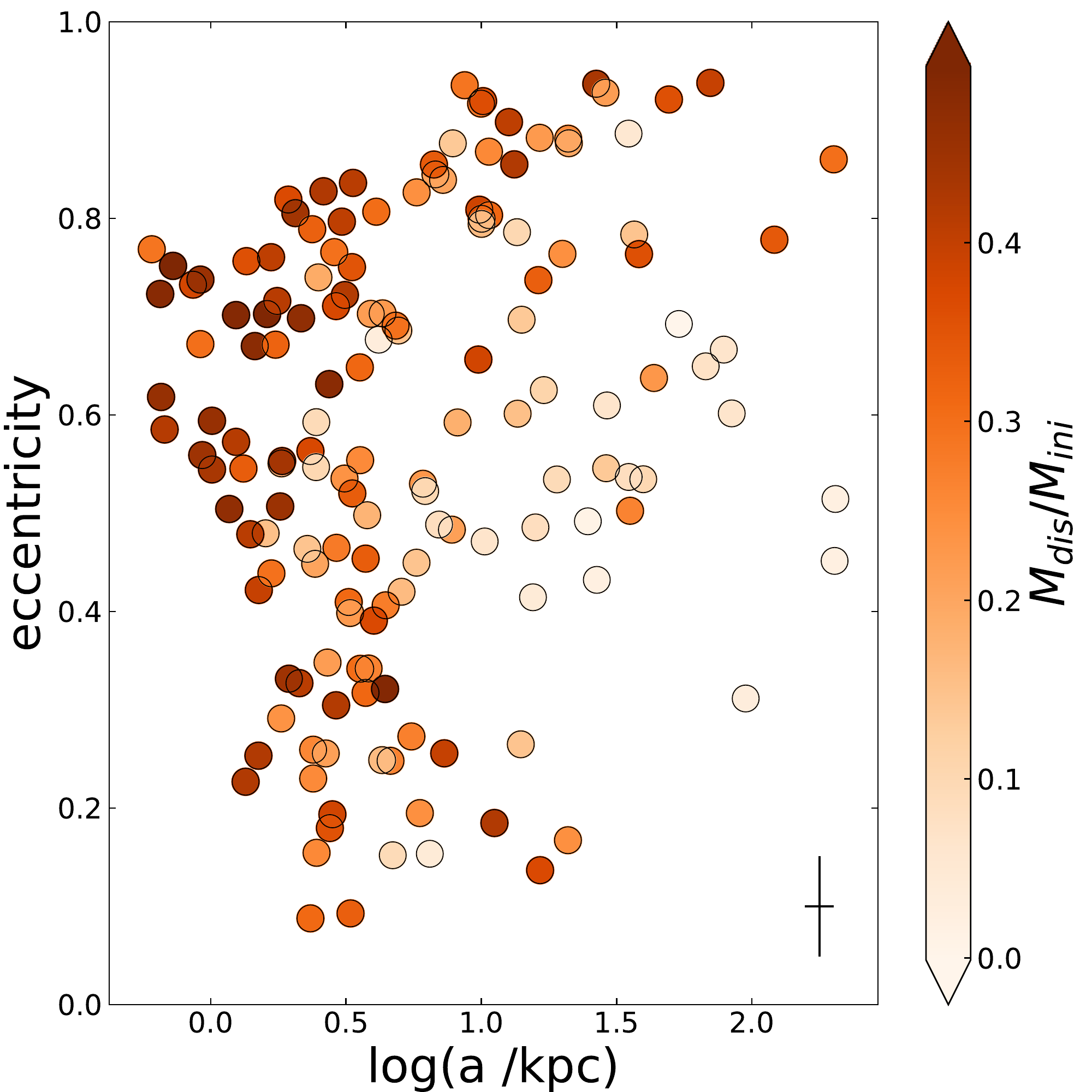}
\caption{Relationship between the eccentricity, the semi-major axis and the
fraction of disrupted mass for the \citet{baumgardtetal2019}'s globular
cluster sample. Typical error bars are  indicated.}
 \label{fig:fig1}
\end{figure}

We begin our analysis of the \citet{bh2018} and \citet{baumgardtetal2019} data with Fig.~\ref{fig:fig1}, which depicts the  relationship between orbital eccentricity, the cluster`s semi-major axis and the fraction of mass lost by disruption. 
Fig.~\ref{fig:fig1} shows that $M_{dis}/M_{ini}$ is a function of both the semi-major axis and the eccentricity.
Likewise, for a given semi-major axis, it is also apparent that the larger the eccentricity of a
cluster's orbit the larger the fraction of mass lost by disruption, as eccentric clusters are brought 
deeper into the Milky Way's potential well. We note that while this result is simply a biproduct of employing \citet{bm2003} to estimate each cluster's initial mass, but it is worth illustrating such a relationship for the actual Milky Way globular cluster system. Highly eccentric clusters are also likely have higher orbital inclinations -- consistent with \citet{piatti2019} --, meaning they should dissolve even quicker than estimates by \citet{baumgardtetal2019} due to disc shocking \citep{webbetal2014}. In fact, \citet{piatti2019}
found that $\sim$ 30 per cent of globular clusters are on orbits with an inclination angle between 20$\degr$ and 50$\degr$. The majority of clusters have even higher inclinations that exceed 50$\degr$. Hence most Galactic globular clusters have undergone several disc crossing compared to those orbiting on more circular orbits in the plane of the disc at a similar distance from the Milky Way centre. Clusters with highly inclined orbits will of course lose more mass \citep{gnedinetal1999, webbetal2014}.

Just as interesting are the regions of Fig.~\ref{fig:fig1} where no globular clusters are observed. In general
Milky Way bulge globular clusters (log($a$ /kpc) $<$ 0.5) with orbits' eccentricities smaller than $\sim$ 0.5 make up a minor
percentage of the total population, as is also the case for outermost ones (log($a$) /kpc) $>$ 1.5 with eccentricities smaller than $\sim$ 0.5. This behaviour would seem to be an intrinsic property of the Milky Way
globular cluster system, however likely for different reasons. Bulge clusters are subjected to a very strong tidal field, and therefore expected to lose mass quickly and will likely dissolve within a Hubble time if they are not extremely massive and compact. Hence the few existing inner region clusters with eccentricities smaller than $\sim$ 0.5 represent the tail end of the initial globular cluster mass and size distributions. 

The lack of outer clusters with low eccentricties is consistent with recent work by \citet{piatti2019}, who studied the kinematics properties of the Milky Way globular cluster system and found that outer globular clusters, independent of the direction of their motions (prograde or retrograde orbits), tend to have more radial orbits than those in the disc of the Milky Way and preferentially have large orbital inclination angles. \citet{piatti2019} also showed that only outer globular clusters that form {\it in-situ} with their host will have radial orbits, accreted globular clusters will have radial orbits regardless of their location in the Milky Way. Finally, by assigning the
same probability to an accreted globular cluster to have a prograde or a retrograde orbit, \citet{piatti2019} found that the accreted to {\it in-situ} globular cluster ratio turns out to be $\sim$ 1. Hence the lack of outer clusters with low eccentricties is likely due to the fact that most outer halo clusters have been accreted via a past merger. So while the lack of inner region clusters with low eccentricities is due to cluster disruption, the lack of outer region clusters with low eccentricities is likely due to no clusters ever forming there in the first place. 

Finally, we also see a region in the top left region of Fig.~\ref{fig:fig1} where no clusters currently exist that extends out to (log($a$) /kpc) $\sim$ 1.0. The lack of clusters in this portion of the diagram can again be attributed to cluster disruption, as a small semi-major axis and high orbital eccentricity would bring a cluster deep into the Galaxy's pontial well. Hence any cluster that formed with this orbit would dissolve quickly. A puzzling population exists however with intermediate eccentricties (0.5 < e < 0.8) and small semi-major axes (log($a$ /kpc) $<$ 0.5), which \citet{baumgardtetal2019} predicts must have been very massive at birth in order to survive a Hubble time. 
While it is entirely possible these are simply some of the most massive clusters every to have formed in the Milky Way, an alternative explanation would be that these clusters did not always have their present day orbit. Inward migration is typically due to either dynamical friction or an accretion event. In the dynamical friction scenario, which \citet{baumgardtetal2019} accounts for, the clusters were born very massive and have had their orbits decay from a weaker tidal field to a stronger one. Hence they have not spent their entire lifetime subjected to weak tides. In the accretion scenario, a cluster initially orbit deep in the potential well of a dwarf and is not stripped until its previous host reaches the inner parts of the Milky Way ans is near dissolution. In this latter scenario the cluster still orbits in a strong tidal field for the majority of its life, but while a member of a dwarf galaxy the tidal field may be compressive \citep{bianchinietal2015,webbetal2017b}.

In order for these GCs to migrate inwards, from a weaker tidal field to the strong one they experience at present day, they were likely subject to 1 of 2 mechanisms. Either they were very massive at birth and strongly affected by dynamical friction (which Baumgardt accounts for). Or they were accreted clusters that orbitted in the inner regions of a dwarf galaxy. Hence it was not until their previous host reached the inner Milky Way and was almost completely disrupted that they were accreted.

\begin{figure}
     \includegraphics[width=\columnwidth]{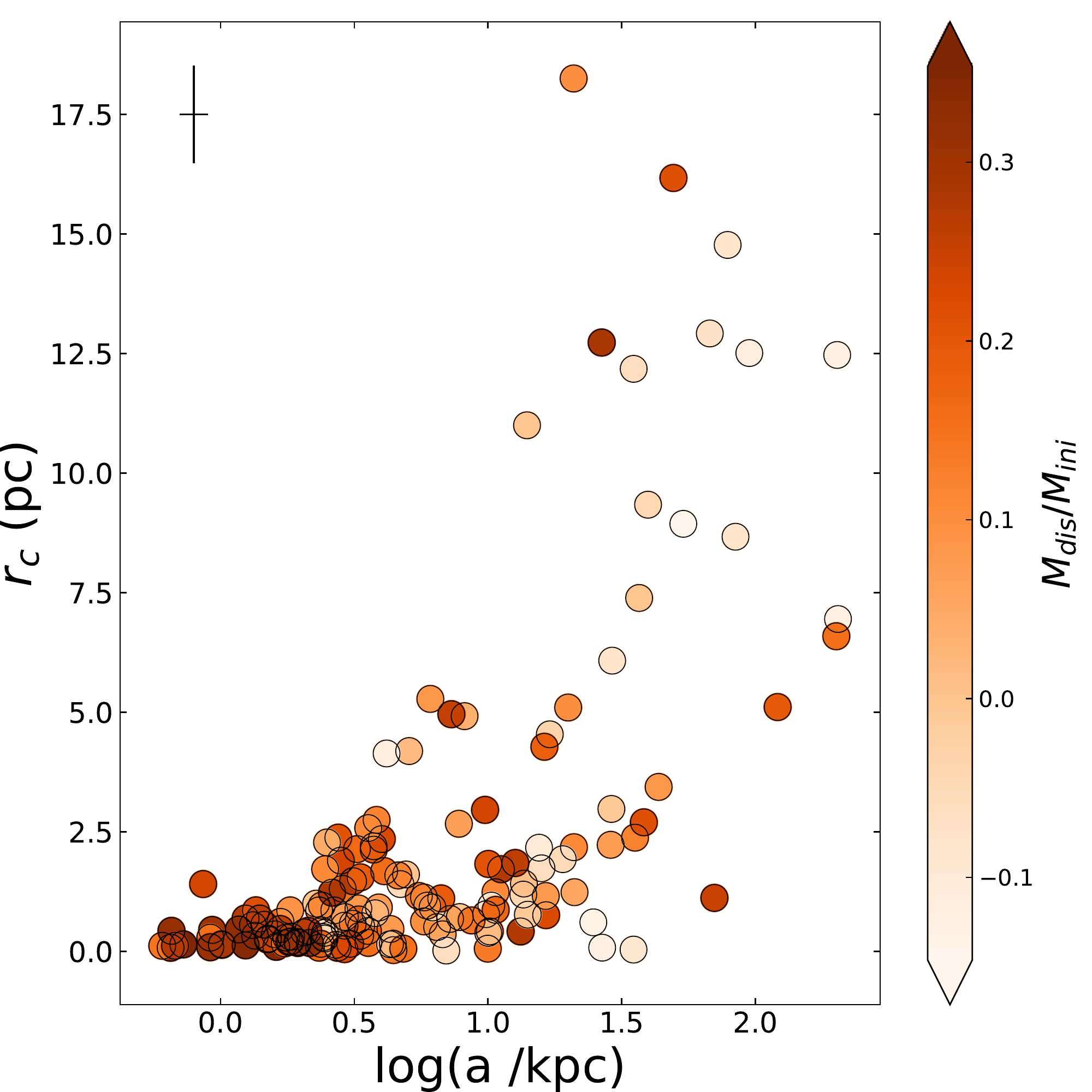}
\caption{Relationship between the core radius, the semi.major axis
and the fraction of mass lost by disruption for the \citet{baumgardtetal2019}'s globular
cluster sample. Typical error bars are  indicated.}
 \label{fig:fig2}
\end{figure}

\begin{figure}
     \includegraphics[width=\columnwidth]{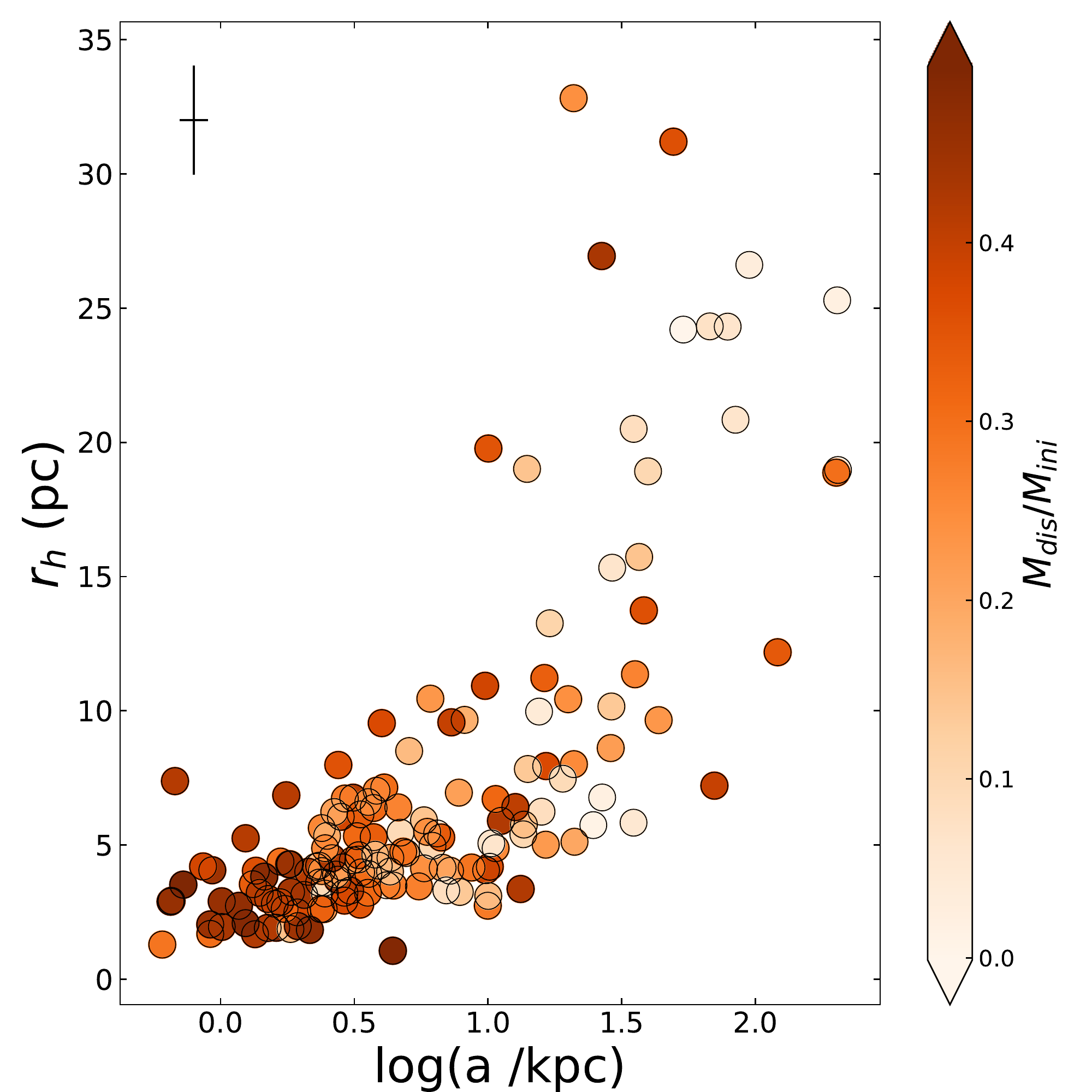}
\caption{Relationship between the half-mass radius, the semi.major axis
and the fraction of mass lost by disruption for the \citet{baumgardtetal2019}'s globular
cluster sample. Typical error bar is indicated.}
 \label{fig:fig3}
\end{figure}

\begin{figure}
     \includegraphics[width=\columnwidth]{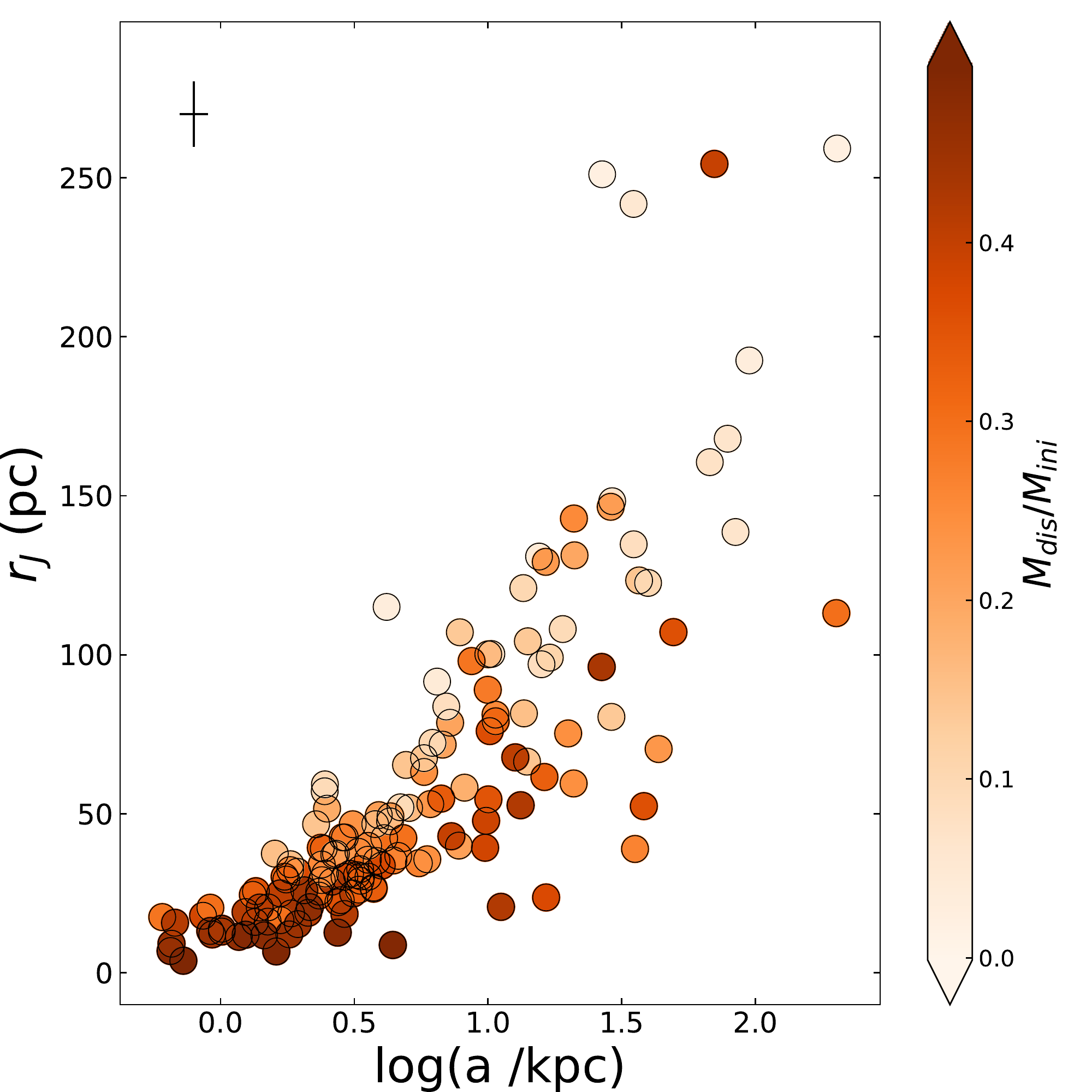}
\caption{Relationship between the Jacobi radius, the semi.major axis
and the fraction of mass lost by disruption for the \citet{baumgardtetal2019}'s globular
cluster sample. Typical error bars are indicated.}
 \label{fig:fig4}
\end{figure}

The different tidal field strengths experienced by globular clusters 
will also affect their structural parameters, and ultimately their
internal dynamical evolutionary stages, with respect to what is
expected while evolving in isolation. Figs.~\ref{fig:fig2}-\ref{fig:fig4} 
illustrate the relationships between $r_c$, $r_{h}$ and $r_J$ with
the semi-major axis and the fraction of mass lost by disruption,
respectively. The three figures highlight a general trend, in the sense
that any of the derived radii increase as a function of the cluster distance 
from the Milky Way centre. This outcome was predicted theoretically
by \citet{hm2010} and \citet{bianchinietal2015}, among others. 
Globular clusters in weaker tidal fields, like those located in the outermost 
regions of the Milky Way can expand naturally, whereas those immersed in
stronger tidal fields (e.g., the Milky Way bulge) do not. Hence we see a wider spread in cluster sizes at larger galactocentric distances than in the inner regions of the Galaxy. We also see an absence of significant outliers in these figures due to accreted clusters, as \citet{miholicsetal2014} has shown the size of an accreted cluster will quickly respond to the tidal field of the Milky Way.

When comparing Figs.~\ref{fig:fig2}-\ref{fig:fig4}, some more subtle differences do
arise. More specifically, $r_c$, $r_{h}$ and $r_J$ all increase with
semi-major axis at different paces and the spread in radius at a given log($a$ /kpc) is also different. Indeed, $r_c$ is the
radius which increases the slowest with log($a$ /kpc), while $r_J$ shows
 the fastest growth. This behaviour is particularly noticeable for globular
clusters inside a circle of 10 kpc (log($a$ /kpc) $<$ 1.0). For instance, the mean
$r_c$, $r_{h}$ and $r_J$ values at log($a$ /kpc) = 1 are 2, 5 and 10 times
those at log($a$ /kpc) = -0.3, respectively. This result expectedly indicates that
the innermost regions of globular clusters are less sensitive to external changes in the Milky Way gravitational field. Similarly, with the cluster's inner regions being less affected by tides than its outer regions, we observe a larger spread in $r_c$ than we do for $r_{h}$ and $r_J$. Since outer region clusters are less affected by the tidal field, with most likely not having expanded to the point of becoming tidally filling \citep{henon1961,ag2013}, the rate of increase in $r_c$, $r_{h}$ and $r_J$ is primarily the result of the initial size and mass distribution of globular clusters and the combined effects of stellar evolution and two-body relaxation. Similar trends are detected in the ancient globular 
clusters of the Large Magellanic Cloud \citep{pm2018}, whose sizes,
\citet{eff87}'s power-law slopes at large radii \citep[$\gamma$,][]{mg04}, ratios of the
cluster radius to Jacobi radius, and the inverse of the concentration
parameter $c$ increase with the deprojected galactocentric distance.
Similar trends are also seen in giant elliptical galaxies 
\citep{harris2009,webbetal2016}.

\citet{hh03} have described the internal dynamics evolution of a star cluster as seen 
in the $r_c/r_h$ versus $r_h/r_J$ plane. As a star cluster expands
to the poing of being tidally filling, it experiences violent relaxation in its core region
followed by two-body relaxation, mass segregation and finally 
core-collapse. These processes result in clusters evolving from having high $r_c/r_h$ and low $r_h/r_J$ to low $r_c/r_h$ and high $r_h/r_J$. We examined such a diagnostic diagram for our 156 globular cluster sample 
and connect the observed trends with those of Figs.~\ref{fig:fig1}-\ref{fig:fig4}
by using the same colour scale given by the range of $M_{dis}/M_{ini}$.
We find that the Milky Way's tidal field has had a role in shaping the internal 
dynamical evolution of clusters as well.

\begin{figure}
     \includegraphics[width=\columnwidth]{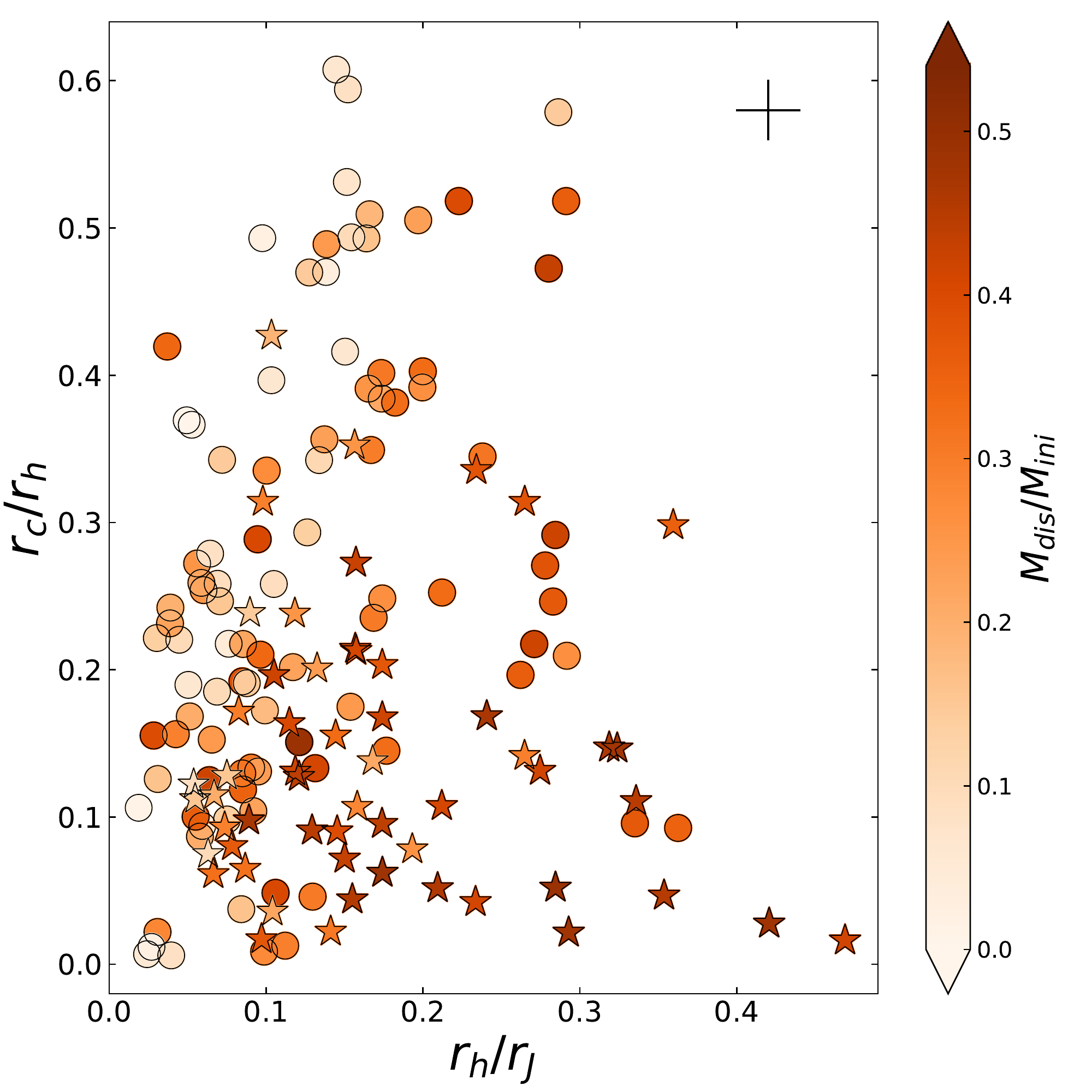}
\caption{Relationship between $r_c$, $r_h$ and $r_J$ radii. Typical error bars are 
indicated.Filled stars and circles represent clusters located inside and outside the
bulge volume (log($a$ /kpc) = 3), respectively.}
 \label{fig:fig5}
\end{figure}

Fig.~\ref{fig:fig5} shows that core-collapse globular clusters are mainly
Milky Way bulge objects (log($a$ / kpc) $\la$ 0.5). They occupy the region
delimited by $r_c/r_h$ $\la$ 0.2 and $r_h/r_J$ $\ga$ 0.30. Likewise, the least
dynamically evolved globular clusters would appear to be the outermost ones ($r_c/r_h$ $\ga$ 0.45).
Although globular clusters have been born with different masses and sizes, and hence
they should stay at different internal dynamics stages even if they were
evolved in isolation, the fact that those at a more 
advanced dynamical stage are the ones located in inner Milky Way regions
reveals that the Milky Way potential well has differentially accelerated their internal 
dynamical evolution. Having lost a large fraction of their initial masses and having their sizes tidally limited will minimize an inner region cluster's relaxation time, allowing it to evolve quickly compared to outer region clusters which lose little mass and can expand to large sizes. Note, for instance, that globular clusters that have lost more
than 45 per cent of their masses by disruption are mostly bulge globular clusters at an advanced evolutionary stage, and conversely, those with $M_{dis}/M_{ini}$ $\la$ 0.10 -- located in the outermost Milky Way 
regions -- are relatively less evolved globular clusters. Some degeneracy exists, of course, as some clusters were likely to have been more with short initial relaxation times and reached the core-collapse phase simply through rapid relaxation. The mostly likely candidates for having their evolution be internally (relaxation) driven as opposed to through interactions with the tidal field are cluster with low $r_c/r_h$ that are not predicted to have lost much mass via tidal stripping (lighter points).

\begin{figure}
     \includegraphics[width=\columnwidth]{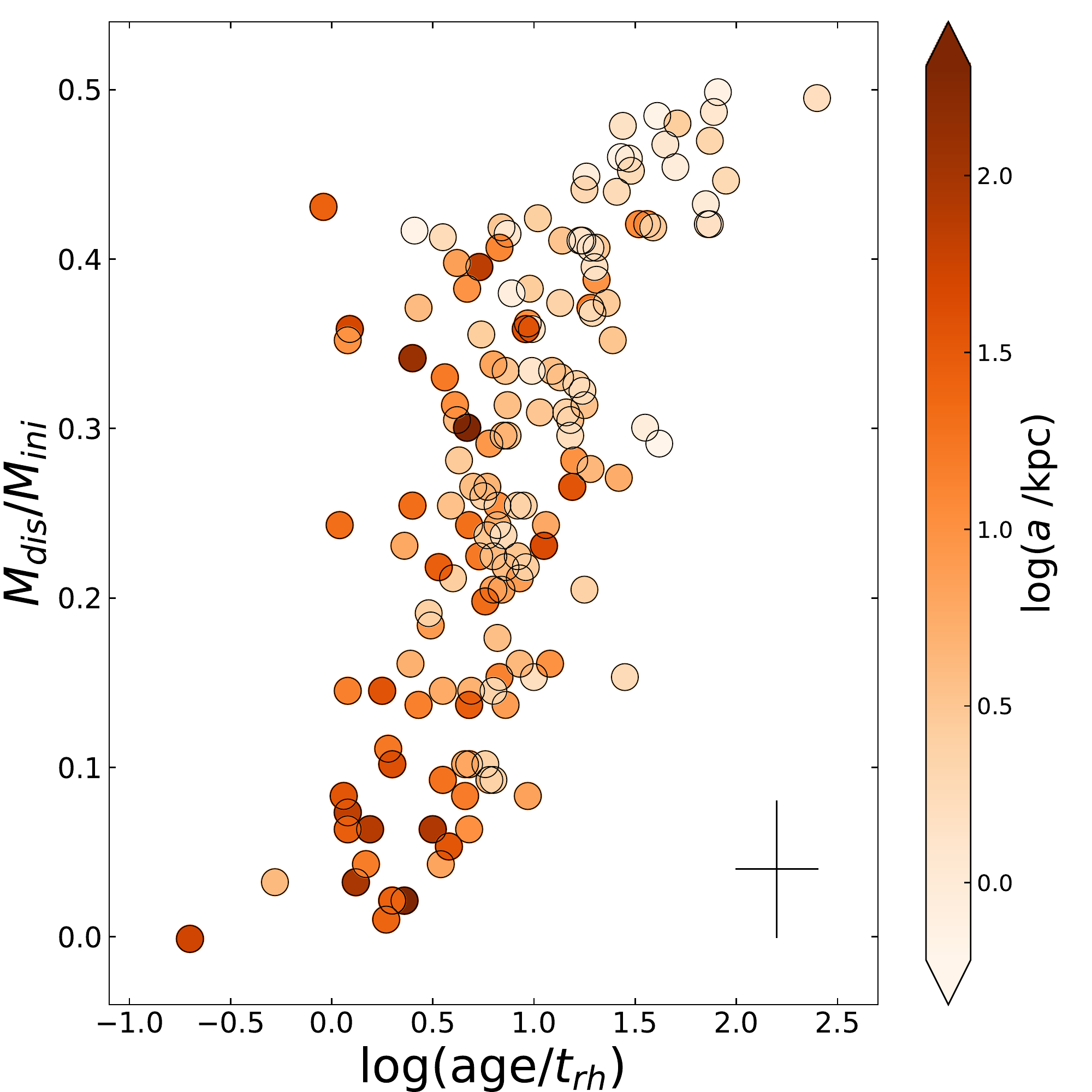}
\caption{Globular cluster fraction of mass disrupted by tidal effects as a
function of the age/$t_{rh}$ ratio.Typical error bars are indicated. }
 \label{fig:fig6}
\end{figure}

Further evidence of such a differential acceleration of the globular cluster
internal dynamics evolution is depicted in Fig.~\ref{fig:fig6}, where we plot
the fraction of mass lost by disruption as a function of the ratio between each cluster's age and its half-mass relaxation time (age/$t_{rh}$). Value for $t_{rh}$ are taken from \citet{baumgardtetal2019}, where $t_{rh}$ is calculated using the formalism of \citet{bh2018}. Globular cluster ages were assumed to be 12$^{-1.5}_{-2.0}$ Gyr 
\citep{kruijssenetal2018}. As can be seen in Fig.~\ref{fig:fig6}, globular clusters with $M_{dis}/M_{ini}$ $\ga$ 0.45 (log($a$ / kpc) $\la$ 0.5) have apparently lived many more median 
relaxation times than their more remote counterparts ($M_{dis}/M_{ini}$ $\la$ 0.10,
which corresponds to log($a / kpc$) $\ga$ 1.5). The small or negligible amount of
mass lost by disruption of the outermost globular clusters lead us to conclude that
most of the amount of mass lost during their lifetimes should have been by
relaxation. Even for a given dynamical age, there is a clear spread in $M_{dis}/M_{ini}$ that can be attributed to cluster orbit (as traced by the each cluster's semi-major axis in the color-bar). This spread helps separate between whether or not a cluster's dynamical age is the result of its formation properties or tidal interactions.

Fig.~\ref{fig:fig6} also shows that globular clusters that have lost between 15 and 40 per cent of their masses by
disruption have similar dynamical ages. These globular
clusters mostly populate the Milky Way disc (0.5 $\la$ log($a$ / kpc) $\la$ 1.4), 
so that the range of mass fraction lost is primarily linked to the globular
cluster's distance from the Galactic centre.
We checked whether globular clusters with highly eccentric orbits could have 
affected their age/$t_{rh}$ ratios because of their many disc crossings, and
found that they do not differentiate from those with smaller eccentricities.
Indeed, Fig.~\ref{fig:fig1} shows that globular clusters with 
eccentricities $>$ 0.8 have lost between $\sim$ 15 and 45
per cent of their masses by disruption, a $M_{dis}/M_{ini}$ range where
log(age/$t_{rh}$) is nearly constant (0.75 $\pm$ 0.25, see Fig.~\ref{fig:fig6}).

\begin{figure*}
	\centering
    \includegraphics[width=0.8\hsize]{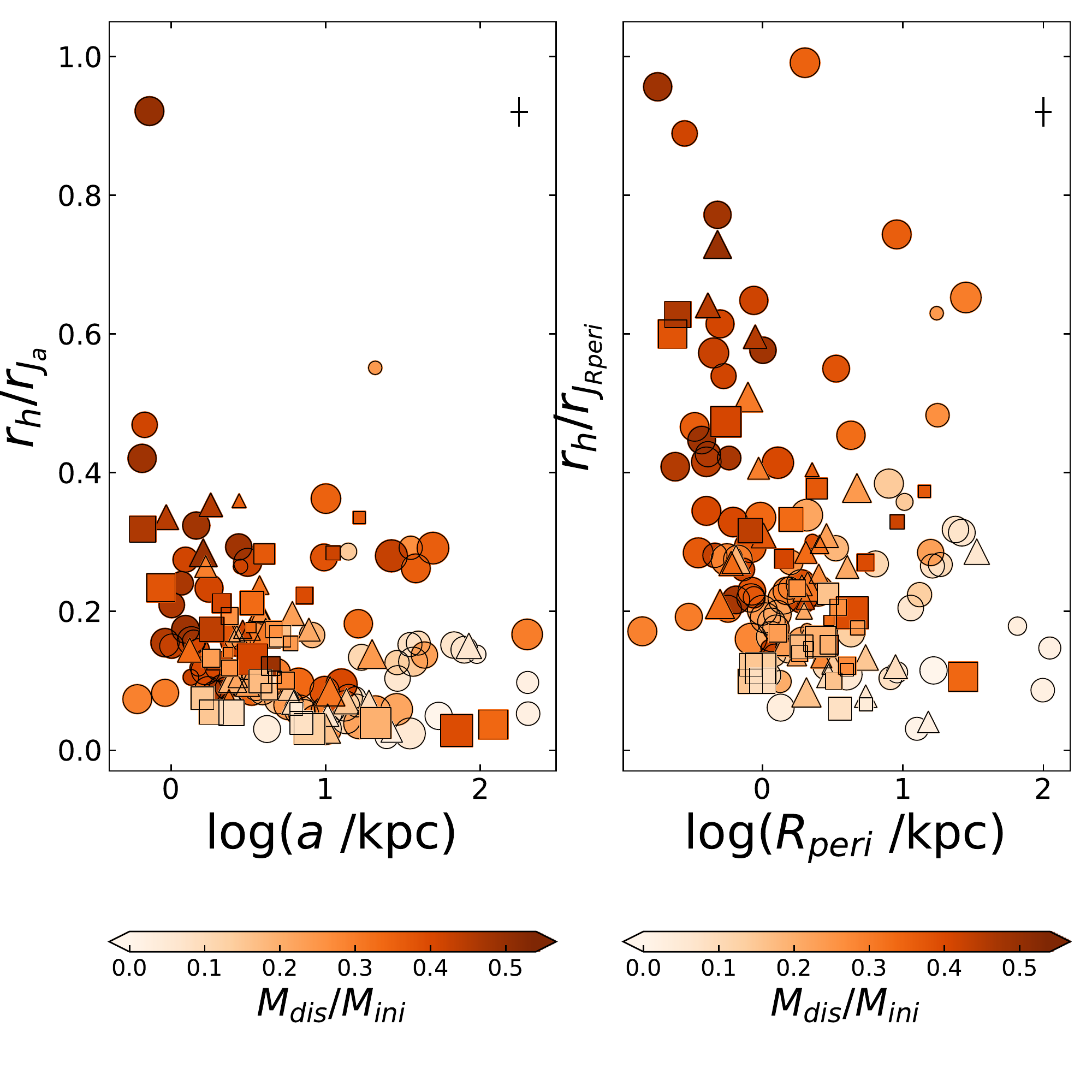}
	\caption{Left Panel: $r_h/r_J$ compared to each cluster's semi-major axis, where $r_J$ is calculated at the cluster's semi-major distance. Right Panel: $r_h/r_J$ compared to each cluster's pericentre, where $r_J$ is calculated at the cluster's perigalactic distance. Typical error bars are indicated. According to  \citet{baumgardtetal2010},
	globular clusters located at Galactocentric distances > 8 kpc and below  $r_h/r_J$ = 0.1 
	or 0.1 $<$ $r_h/r_J$ $<$ 0.3 are genuine compact and extended objects
	at birth, respectively. Symbol sizes are proportional to the orbit eccentricities
	(0 $<$ $\epsilon$ $<$ 1), while triangles, squares and circles correspond to
	orbit inclinations ($i$) between $i$ $\le$ 30$\degr$, 30$\degr$ $<$ $i$ $\le$ 60$\degr$
	and $i$ $>$ 60$\degr$, respectively.}
 	\label{fig:fig7}
\end{figure*}

\citet{baumgardtetal2010} investigated the $r_h/r_J$ ratio of Galactic globular clusters as given by 
the Milky Way' s potential and found that the $r_h/r_J$ ratio is a good discriminator
between globular clusters that are compact and extended at birth. This ratio has also long been the criteria for determining whether a cluster is tidally filling or not, with clusters that have $r_h/r_J > 0.145$ considered to be tidally filling \citep{h1965}.
\citet{baumgardtetal2010} found  that
for Galactocentric distances larger than 8 kpc, $r_h/r_J$ ratios smaller than
0.05 correspond to initially compact objects, while $r_h/r_J$ ratios between 0.1 and
0.3 point to globular clusters intrinsically larger, and not because of expansion in a weaker tidal field. Both groups of globular clusters would have an {\it in-situ} origin. To further expand their study on compact and extended clusters, in the left panel of Fig.~\ref{fig:fig7} we reproduce \citet{baumgardtetal2010}'s figure 2 comparing $r_h/r_J$ to each cluster's semi-major axis for our enlarged
globular cluster sample.  
In order to
make Fig.~\ref{fig:fig7} as informative as possible, we included some other cluster orbital motion
parameters, namely, the orbit eccentricity and inclination; the latter split in arbitrary bins. In both panels, symbol sizes are proportional to each cluster's orbital eccentricity and symbol shapes are related to their orbital inclinations ($i$). For comparison purposes we also show in the right panel $r_h/r_J$ versus easch cluster's perigalactic distance, where in this case $r_J$ is calculated at perigalactic.


For reference purposes, we include Table~\ref{tab:table1} in the Appendix containing all the relevant informatin used in making Figure \ref{fig:fig7}. Table~\ref{tab:table1} serves as a useful source for identifying clusters that are tidally filling or under-filling both on average and near pericentre. Clusters that are tidally under-filling at pericentre have likely had their evolution governed by two-body relaxation, while clusters that are on average tidally filling have likely had their evolution strongly shaped by the tidal field. For all other clusters, a comibnation of relaxation and tidal interactions have played a role in their evolution and its not neccesarily the case that one mechanism is more dominant over the other.

The left panel of Fig.~\ref{fig:fig7} shows that, in general, the smaller the distance from 
the Milky Way centre the larger the $r_h/r_J$ ratio. This would seem to be a consequence 
of inner region clusters easily expanding to the point of being tidally filling due to stellar evolution and two-body relaxation. Hence they have experienced a significant amount of mass loss due to disruption, primarily of low-mass stars that have segregated outwards, leaving the surviving globular clusters with their more massive members \citep{bm2003,khalisietal2007}. Therefore as expected, their $M_{dis}/M_{ini}$ ratios are amongst the largest in the dataset. In the inner regions of the galaxy there is significant scatter about this general trend, as first pointed out by \citet{baumgardtetal2010}, which is likely a remnant of the range of inititial masses and sizes clusters can have at birth.

In the outer regions of the galaxy (beyond $\sim$ 8 kpc), tidally under-filling clusters continue to follow the general trend of $r_h/r_J$ with galactocentric distance down to a mean value of 0.05. Given the relatively weak external tidal field that these clusters are subject to, their evolution is almost entirely dominated by internal processes. However, beyond 8 kpc there also exists a population of tidally filling clusters. Previous work by \citet{baumgardtetal2010} also recognized that outer region clusters with log($a$ / kpc) $>$ 0.9 could be separated into tidally under-filling ($r_h/r_J$ $<$ 0.05) and filling (0.1 $<$ $r_h/r_J$ $<$ 0.3) populations. We further separate the filling clusters into two additional sub-populations. with $r_h/r_J \sim$ 0.2 (barely tidally filling) and $r_h/r_J >=$ 0.3 (tidally over-filling), based on an apparent gap in the $r_h/r_J$ distribution of outer region clusters. Given the relatively weak tidal field the clusters experience, tidally filling outer region clusters are likely to either be accreted, have high eccentricities, or formed quite extended.

Independent of semi-major axis, we identify a potential third group of globular clusters with 0.4 $<$ $r_h/r_J$ $<$ 1.0. Clusters in this group also have highly eccentric and inclined orbits, such that they are subject to strong tidal shocks at pericentre and during disc passages. Given that these four clusters are severly tidally over-filling, they should be in the process of dissolving.

The right panel of Fig.~\ref{fig:fig7} is also of interest, as it indicates that almost every cluster is tidally filling at perigalacticon. Hence, while the few that are not tidally filling at perigalacticon can be considered to have had their evolution be relaxation dominated, the majority of clusters are affected by the tidal field to some degree. Clusters that are tidally filling at their semi-major axis (left panel) are most likely to have their evolution be tidally dominated.


\section{Conclusions}

We have analysed the relationships between the Milky Way's globular 
clusters structural parameters, their internal dynamical stages, and the the fraction of mass lost by tidal effects (disruption). There has long been
a general understanding that the Milky Way gravitational field has played a strong role   
in the globular cluster mass loss process along their lifetimes.

We made use of publicly available positions, space velocity components, orbital motion parameters, core, half-mass and tidal radii, relaxation times, current and 
initial  masses of 156 Milky Way globular clusters, from which we addressed the
 issue about at what extend the Milky Way potential well has acted in shaping
 their present-day sizes and dynamical age. As suggested by
previous theoretical results, we started by analysing the relationship between the fraction of mass lost by disruption, the globular cluster distance to the Milky Way centre and the eccentricity of their orbits. We find an absence of cluster with short semi-major axes and either low or high orbital eccentricities, likely due to the fact that any clusters born in this regime would reach dissolution fairly quickly. Although a puzzling population of clusters with intermediate eccentricties and short semi-major axes is observed, also with relatively high orbital inclinations. There is also a lack of outer region clusters with low orbital eccentricities, likely due to outer region clusters primarily being accreted clusters that have orbits that are comparable to the radial orbit in which their progenitor host galaxy fell in on.

The core, half-mass and Jacobi radii of Galactic clusters show different rates of increase with Galactocentric distance, in the sense that the core radii increases 
slower than the half-mass radii, which in turn increases at a
slower pace than Jacobi radii. This outcome would seem to suggest that
the inner most regions of globular clusters are less
sensitive to changes in the Milky Way potential with the Galactocentric
distance. On the contrary, their outermost regions would appear more
vulnerable.

From the different fractions of mass lost by disruption of each cluster, we found that the Milky Way's gravitational field has differentially accelerated the internal dynamical evolution of its globular clusters. Having been stripped of a large fraction of their initial mass, and having their sizes tidally limited, inner region clusters will have shorter relaxatin times than outer region clusters. Hence globular clusters located in the bulge would seem to be at a more advanced internal dynamical evolutionary stage, approaching to or at the core-collapse stage, while more distant globular clusters have lived many fewer median relaxation times than their innermost counterparts.

Finally, we confirmed the existence of two intrinsically different size globular
cluster groups as identified by \citet{baumgardtetal2010} at large Galactocentric
distances. That is to say we find globular clusters can primarily be split into groups with $r_h/r_J$ $\sim$ 0.05 and 0.20 that have lost a very small fraction of their initial mass. However, we point out that interspersed within both groups are clusters that have lost a large percentage of their masses by disruption. Hence it is more difficult to distinguish between clusters that are compact and extended at birth from their $r_h/r_J$ ratio alone. There is a third group of objects with  $r_h/r_J$ $>$ 0.4 which have lost even more
mass due to tidal effects. In both cases, clusters that have lost a large portion of their initial mass have highly eccentric orbits with large inclination angles. which leads to them experiencing more tidal shocks at perigalcticon and while crossing the
disc. Hence they have lost an additional fraction of mass over what their semi-major axis alone implies. These clusters are, most likely, clusters that have been accreted by the Milky Way and have had their structural properties respond to their new host.

\section*{Acknowledgements}
We thank the referee, Florent Renaud, for a timely and constructive report to improve the manuscript.
We thank Henny J.G.L.M. Lamers and Mark Gieles for their reading through an earlier 
version of this manuscript and timely comments and Holger Baumgardt for providing 
us with his globular cluster database.










\appendix

\begin{table*}
\caption{Milky Way globular clusters' parameters used to build Fig.~\ref{fig:fig7}.}
\label{tab:table1}
\begin{tabular}{@{}lcccccc}\hline

ID  & $a$      &  $R_{peri}$    & $rh/r_{J_{a}}$ & $rh/r_{J_{Rperi}}$ & $\epsilon$ & $i$ \\
     & (kpc)   &  (pc)          &                    &     &   & (deg)                  \\\hline
         NGC104 &   6.45$\pm$ 0.01   &  5.46$\pm$ 0.01 & 0.06 &0.07& 0.15$\pm$0.00  & 27.78$\pm$ 0.80 \\
         NGC288  &   8.17$\pm$ 0.28   &  3.33$\pm$ 0.49&  0.17& 0.29& 0.59$\pm$0.05 & 124.77$\pm$ 1.83 \\
         NGC362 &    6.76$\pm$ 0.23   &  1.05$\pm$ 0.21&  0.06& 0.19& 0.84$\pm$0.03 &  85.40$\pm$ 3.63 \\
       Whiting1 &   35.46$\pm$ 4.48   & 17.64$\pm$ 4.08&  0.29& 0.48& 0.50$\pm$0.10 &  71.76$\pm$ 4.24 \\
        NGC1261 &   10.67$\pm$ 0.85   &  1.41$\pm$ 0.36&  0.06& 0.22& 0.87$\pm$0.03 & 110.52$\pm$ 9.87 \\
           Pal1 &   16.49$\pm$ 0.89   & 14.23$\pm$ 0.91 & 0.34& 0.37& 0.14$\pm$0.05 &  14.45$\pm$ 5.28 \\
            AM1 &  203.59$\pm$89.87   & 98.84$\pm$39.38& 0.05 &0.09& 0.51$\pm$0.26 & 110.20$\pm$41.25 \\
       Eridanus &   84.24$\pm$14.67  &  33.56$\pm$23.03 &0.15 &0.29& 0.60$\pm$0.22 &  46.82$\pm$28.63 \\
           Pal2 &   20.95$\pm$ 1.83  &   2.49$\pm$ 1.80 &0.06 &0.23& 0.88$\pm$0.08 & 144.05$\pm$35.09 \\
        NGC1851 &    9.98$\pm$ 0.11   &  0.83$\pm$ 0.05 &0.03 &0.16& 0.92$\pm$0.00 &  93.82$\pm$ 1.71 \\
        NGC1904 &   10.15$\pm$ 0.59   &  0.82$\pm$ 0.33 &0.06 &0.29& 0.92$\pm$0.03 &  81.65$\pm$ 5.22 \\
        NGC2298 &    9.84$\pm$ 0.73   &  1.88$\pm$ 0.79 &0.08 &0.24& 0.81$\pm$0.07 & 121.07$\pm$ 5.94 \\
        NGC2419  &  53.74$\pm$ 4.06   & 16.52$\pm$ 2.68 &0.05 &0.11& 0.69$\pm$0.05 &  60.48$\pm$ 6.22 \\
          Pyxis  &  78.74$\pm$15.69   & 26.26$\pm$ 8.98& 0.14 &0.31& 0.67$\pm$0.11 & 102.92$\pm$ 3.94 \\
        NGC2808  &   7.85$\pm$ 0.06   &  0.97$\pm$ 0.02& 0.03 &0.12& 0.88$\pm$0.00 &  13.05$\pm$20.09 \\
             E3  &  11.18$\pm$ 1.40   &  9.11$\pm$ 0.44& 0.28 &0.33& 0.18$\pm$0.10 &  28.91$\pm$ 9.65 \\
           Pal3  &  94.87$\pm$25.51   & 65.31$\pm$23.15& 0.14 &0.18& 0.31$\pm$0.23 &  71.59$\pm$15.90 \\
        NGC3201  &  15.85$\pm$ 0.61   &  8.15$\pm$ 0.09 &0.06 &0.10& 0.49$\pm$0.02 & 152.31$\pm$ 2.73 \\
           Pal4 &   67.51$\pm$14.82  &  23.66$\pm$22.78& 0.15 &0.32& 0.65$\pm$0.28 &  67.95$\pm$55.01 \\
         Crater &  202.23$\pm$89.57  & 110.92$\pm$47.45& 0.10 &0.15& 0.45$\pm$0.29 & 109.03$\pm$60.62 \\
        NGC4147 &   13.24$\pm$ 1.32   &  1.92$\pm$ 0.77& 0.06 &0.22& 0.86$\pm$0.06 &  83.73$\pm$ 2.13 \\
        NGC4372  &   5.07$\pm$ 0.08   &  2.94$\pm$ 0.14& 0.16 &0.23& 0.42$\pm$0.02 &  28.59$\pm$ 4.93 \\
         Rup106 &   19.94$\pm$ 2.04   &  4.71$\pm$ 0.67& 0.14 &0.38& 0.76$\pm$0.04 &  46.00$\pm$ 8.58 \\
        NGC4590  &  19.03$\pm$ 1.44   &  8.86$\pm$ 0.42& 0.07 &0.12& 0.53$\pm$0.04 &  41.04$\pm$ 8.30 \\
        NGC4833  &   4.09$\pm$ 0.16   &  0.79$\pm$ 0.12& 0.17 &0.51& 0.81$\pm$0.03 &  44.23$\pm$ 9.88 \\
        NGC5024  &  15.52$\pm$ 0.92   &  9.09$\pm$ 0.16& 0.08 &0.11& 0.41$\pm$0.04 &  74.80$\pm$ 1.44 \\
        NGC5053  &  13.99$\pm$ 0.68   & 10.28$\pm$ 0.09& 0.29 &0.36& 0.26$\pm$0.04 &  76.11$\pm$ 1.15 \\
        NGC5139  &   4.18$\pm$ 0.03   &  1.35$\pm$ 0.04 &0.03 &0.06& 0.68$\pm$0.01 & 138.06$\pm$ 0.89 \\
        NGC5272  &  10.29$\pm$ 0.21   &  5.44$\pm$ 0.09 &0.05 &0.08& 0.47$\pm$0.01 &  56.37$\pm$ 2.46 \\
        NGC5286  &   7.21$\pm$ 0.45   &  1.16$\pm$ 0.24 &0.05 &0.16& 0.84$\pm$0.03 & 125.18$\pm$18.04 \\
            AM4  & 200.68$\pm$92.69   & 28.07$\pm$ 2.24& 0.17 &0.65& 0.86$\pm$0.07 &  83.67$\pm$10.52 \\
        NGC5466  &  36.74$\pm$13.50   &  7.95$\pm$ 2.63& 0.13 &0.38& 0.78$\pm$0.10 & 107.25$\pm$ 2.20 \\
        NGC5634  &  14.09$\pm$ 1.86  &   4.27$\pm$ 2.30& 0.08 &0.17& 0.70$\pm$0.14 &  64.17$\pm$ 2.87 \\
        NGC5694  &  35.01$\pm$ 3.92  &   3.98$\pm$ 0.95 &0.02 &0.11& 0.89$\pm$0.03 & 124.95$\pm$ 8.69 \\
         IC4499  &  17.02$\pm$ 1.83   &  6.38$\pm$ 1.24 &0.13 &0.27& 0.63$\pm$0.07 & 112.45$\pm$ 1.86 \\
        NGC5824  &  26.72$\pm$ 5.64   & 15.17$\pm$ 5.45 &0.03 &0.04& 0.43$\pm$0.18 &  57.22$\pm$ 2.76 \\
           Pal5  &  20.90$\pm$ 4.11   & 17.40$\pm$ 6.04 &0.55 &0.63& 0.17$\pm$0.20 &  65.13$\pm$ 2.14 \\
        NGC5897  &   6.09$\pm$ 0.87  &   2.86$\pm$ 1.05 &0.20 &0.31& 0.53$\pm$0.14 &  59.50$\pm$12.47 \\
        NGC5904  &  13.55$\pm$ 0.53   &  2.90$\pm$ 0.05 &0.04 &0.12& 0.79$\pm$0.01 &  74.09$\pm$ 0.66  \\
        NGC5927  &   4.70$\pm$ 0.09   &  3.99$\pm$ 0.02 &0.10 &0.12& 0.15$\pm$0.02 &   9.13$\pm$20.45 \\
        NGC5946  &   3.33$\pm$ 0.46   &  0.83$\pm$ 0.43 &0.09 &0.22& 0.75$\pm$0.12 &  77.17$\pm$ 1.91 \\
        NGC5986   &  2.86$\pm$ 0.28   &  0.67$\pm$ 0.31 &0.10 &0.27&0.77$\pm$0.10  & 60.88$\pm$11.67 \\
        FSR1716   &  3.74$\pm$ 0.31   &  2.55$\pm$ 0.42 &0.24 &0.30& 0.32$\pm$0.09 &  32.33$\pm$11.77 \\
          Pal14  &  49.36$\pm$ 4.57  &   3.90$\pm$ 1.90& 0.29 &1.70& 0.92$\pm$0.04 &  75.06$\pm$193.30 \\
         Lynga7  &   3.24$\pm$ 0.26   &  1.91$\pm$ 0.39& 0.17 &0.24& 0.41$\pm$0.09 &  36.06$\pm$ 9.71 \\
        NGC6093  &   1.93$\pm$ 0.14  &   0.35$\pm$ 0.10& 0.08 &0.28& 0.82$\pm$0.05 &  97.02$\pm$ 2.65 \\
        NGC6121  &   3.36$\pm$ 0.06   &  0.55$\pm$ 0.08 &0.13 &0.47& 0.84$\pm$0.02 &   4.96$\pm$ 7.66 \\
        NGC6101  &  29.13$\pm$ 5.14   & 11.37$\pm$ 1.07& 0.10 &0.20& 0.61$\pm$0.07 & 143.36$\pm$ 8.63 \\
        NGC6144  &   2.82$\pm$ 0.30   &  2.27$\pm$ 0.31& 0.26 &0.30& 0.19$\pm$0.10 &  116.71$\pm$ 3.09 \\
        NGC6139  &   2.43$\pm$ 0.36  &   1.34$\pm$ 0.45 &0.07 &0.10& 0.45$\pm$0.15 &  62.30$\pm$ 5.09 \\
           Ter3  &   2.76$\pm$ 0.23   &  2.26$\pm$ 0.11& 0.36 &0.40& 0.18$\pm$0.07 &  42.91$\pm$ 6.72 \\
        NGC6171  &   2.34$\pm$ 0.09  &  1.02$\pm$ 0.15& 0.17 &0.31& 0.56$\pm$0.05  & 49.25$\pm$ 6.55 \\
    ESO452-SC11  &   1.62$\pm$ 0.19  &  0.48$\pm$ 0.18& 0.28 &0.73& 0.70$\pm$0.10  & 48.30$\pm$ 1.71 \\
        NGC6205  &   4.94$\pm$ 0.08  &  1.55$\pm$ 0.04& 0.07 &0.15& 0.69$\pm$0.01 & 105.00$\pm$ 0.74 \\
        NGC6229  &  16.44$\pm$ 1.73  &  1.94$\pm$ 1.49& 0.04 &0.16& 0.88$\pm$0.09 &  73.83$\pm$16.27 \\
        NGC6218   &  3.57$\pm$ 0.06  &  2.35$\pm$ 0.10& 0.13 &0.17& 0.34$\pm$0.02 &  36.79$\pm$ 4.40 \\
        FSR1735  &   3.13$\pm$ 0.44  &  0.87$\pm$ 0.27& 0.27 &0.65& 0.72$\pm$0.08 &  67.41$\pm$ 5.27 \\
        NGC6235  &  13.62$\pm$ 9.74  &  5.43$\pm$ 1.53& 0.07 &0.13& 0.60$\pm$0.30 &  39.53$\pm$ 2.06 \\
        NGC6254  &   3.28$\pm$ 0.05  &  1.97$\pm$ 0.04& 0.12 &0.16& 0.40$\pm$0.01 &  42.78$\pm$ 1.89 \\
        NGC6256  &   2.34$\pm$ 0.32  &  2.13$\pm$ 0.47& 0.14 &0.15& 0.09$\pm$0.14 &  18.53$\pm$ 8.09 \\
          Pal15  &  26.61$\pm$ 2.86  &  1.68$\pm$ 0.92& 0.28 &1.75& 0.94$\pm$0.03 & 110.49$\pm$39.89 \\
        NGC6266   &  1.59$\pm$ 0.06  &  0.83$\pm$ 0.08& 0.07 &0.12& 0.48$\pm$0.04 &  29.05$\pm$ 6.51 \\
        NGC6273   &  2.28$\pm$ 0.14  &  1.22$\pm$ 0.11& 0.09 &0.14& 0.46$\pm$0.05 &  99.61$\pm$12.71 \\
        NGC6284   &  4.32$\pm$ 0.62  &  1.28$\pm$ 0.39& 0.09 &0.20& 0.70$\pm$0.09 &  90.47$\pm$ 0.62 \\
\hline
\end{tabular}
\end{table*}

\setcounter{table}{0}
\begin{table*}
\caption{\it continued.}
\label{tab:table1}
\begin{tabular}{@{}lcccccc}\hline

ID  & $a$      &  $R_{peri}$    & $rh/r_{J_{a}}$ & $rh/r_{J_{Rperi}}$ & $\epsilon$ & $i$ \\
     & (kpc)   &  (pc)          &                    &     &   & (deg)                  \\\hline
        NGC6287   &  3.56$\pm$ 0.80  &  1.25$\pm$ 0.26 &0.09 &0.18& 0.65$\pm$0.10 &  95.88$\pm$ 0.50 \\
        NGC6293   &  1.76$\pm$ 0.46  &  0.50$\pm$ 0.18 &0.23 &0.61& 0.72$\pm$0.11 & 130.42$\pm$54.17 \\
        NGC6304   &  2.39$\pm$ 0.24 &   1.77$\pm$ 0.23& 0.19 &0.23& 0.26$\pm$0.09 &  19.66$\pm$31.83 \\
        NGC6316   &  3.12$\pm$ 0.89  &  1.45$\pm$ 0.87& 0.09 &0.15& 0.54$\pm$0.24 &  34.75$\pm$ 2.86 \\
        NGC6341   &  5.76$\pm$ 0.12  &  1.00$\pm$ 0.09& 0.07 &0.20& 0.83$\pm$0.01 &  83.79$\pm$ 1.45 \\
        NGC6325   &  1.35$\pm$ 0.34  &  1.04$\pm$ 0.13& 0.11 &0.13& 0.23$\pm$0.20 & 106.50$\pm$ 1.38 \\
        NGC6333   &  3.90$\pm$ 0.39  &  1.16$\pm$ 0.15& 0.09& 0.19& 0.70$\pm$0.04 &  64.73$\pm$ 3.63 \\
        NGC6342   &  1.50$\pm$ 0.25  &  1.12$\pm$ 0.34 &0.12& 0.15& 0.25$\pm$0.17 &  62.88$\pm$ 3.56 \\
        NGC6356   &  5.76$\pm$ 1.16  &  3.17$\pm$ 1.58& 0.09& 0.13& 0.45$\pm$0.21 &  43.18$\pm$ 2.20 \\
        NGC6355   &  1.50$\pm$ 0.46  &  0.87$\pm$ 0.24 &0.15& 0.22& 0.42$\pm$0.20 & 103.08$\pm$ 0.84 \\
        NGC6352   &  3.28$\pm$ 0.30  &  2.98$\pm$ 0.25 &0.18& 0.19& 0.09$\pm$0.09 &  13.33$\pm$13.74 \\
         IC1257   & 10.03$\pm$ 1.37  &  2.01$\pm$ 0.72& 0.36& 0.99& 0.80$\pm$0.07 & 158.42$\pm$31.57 \\
           Ter2  &   0.65$\pm$ 0.23  &  0.18$\pm$ 0.06& 0.42& 0.96& 0.72$\pm$0.13 & 158.97$\pm$26.47 \\
        NGC6366  &   3.74$\pm$ 0.10  &  2.04$\pm$ 0.11& 0.20& 0.29& 0.45$\pm$0.02 &  34.87$\pm$ 4.76 \\
           Ter4   &  0.93$\pm$ 0.29  &  0.41$\pm$ 0.22& 0.34& 0.64& 0.56$\pm$0.22 &  46.85$\pm$ 2.07 \\
            HP1   &  1.24$\pm$ 0.33  &  0.53$\pm$ 0.23& 0.27& 0.54& 0.57$\pm$0.18 &  87.68$\pm$ 0.54 \\
        NGC6362   &  3.83$\pm$ 0.05  &  2.52$\pm$ 0.09 &0.20& 0.25& 0.34$\pm$0.02 &  44.18$\pm$ 3.07 \\
           Lil1   &  0.61$\pm$ 0.16  &  0.14$\pm$ 0.11& 0.07& 0.17& 0.77$\pm$0.17 & 162.65$\pm$31.56 \\
        NGC6380   &  1.35$\pm$ 0.37  &  0.33$\pm$ 0.10& 0.16& 0.47& 0.76$\pm$0.09 & 157.67$\pm$14.22 \\
           Ter1   &  0.86$\pm$ 0.31  &  0.23$\pm$ 0.17 &0.23& 0.60& 0.73$\pm$0.19 &  13.39$\pm$11.30 \\
           Ton2  &   2.90$\pm$ 0.44  &  2.02$\pm$ 0.31& 0.17& 0.22& 0.30$\pm$0.12 &  35.55$\pm$ 3.74 \\
        NGC6388  &   2.45$\pm$ 0.05  &  1.11$\pm$ 0.02& 0.06& 0.11& 0.55$\pm$0.01 & 154.64$\pm$ 5.03 \\
        NGC6402  &   2.50$\pm$ 0.13  &  0.65$\pm$ 0.17& 0.10& 0.28& 0.74$\pm$0.06 &  46.95$\pm$ 5.07 \\
        NGC6401  &   1.32$\pm$ 0.38  &  0.60$\pm$ 0.41& 0.14& 0.27& 0.55$\pm$0.26 &  51.33$\pm$ 6.60 \\
        NGC6397   &  4.43$\pm$ 0.02  &  2.63$\pm$ 0.03& 0.10& 0.13& 0.41$\pm$0.00 &  47.06$\pm$ 0.42 \\
           Pal6   &  2.06$\pm$ 0.37  &  0.40$\pm$ 0.10& 0.12& 0.42& 0.81$\pm$0.06 &  82.84$\pm$ 0.78 \\
        NGC6426  & 121.07$\pm$83.20  & 26.84$\pm$ 5.46& 0.04& 0.11& 0.78$\pm$0.16 &  20.93$\pm$ 4.24 \\
          Djor1   & 70.25$\pm$71.73  &  4.36$\pm$ 1.76 &0.03& 0.20& 0.94$\pm$0.07 &   6.00$\pm$17.61 \\
           Ter5   &  1.82$\pm$ 0.31  &  0.82$\pm$ 0.32& 0.05& 0.10& 0.55$\pm$0.15 &  12.69$\pm$74.47 \\
        NGC6440   &  0.91$\pm$ 0.19  &  0.30$\pm$ 0.12 &0.08& 0.19& 0.67$\pm$0.13 & 113.79$\pm$ 2.79 \\
        NGC6441  &   2.45$\pm$ 0.08  &  1.00$\pm$ 0.08 &0.05& 0.10& 0.59$\pm$0.03 &  20.82$\pm$ 8.51 \\
           Ter6  &   0.92$\pm$ 0.28  &  0.24$\pm$ 0.09 &0.15& 0.41& 0.74$\pm$0.12 & 157.28$\pm$21.98 \\
        NGC6453  &   2.91$\pm$ 0.84  &   1.56$\pm$ 0.90& 0.16& 0.23& 0.46$\pm$0.26 &  78.02$\pm$ 1.02 \\
           UKS1  &   0.65$\pm$ 0.19  &   0.25$\pm$ 0.22& 0.32& 0.63& 0.62$\pm$0.29 &  11.86$\pm$37.46 \\
        NGC6496  &   7.78$\pm$ 3.87  &   4.02$\pm$ 0.94& 0.17& 0.26& 0.48$\pm$0.27 &  30.99$\pm$ 5.88 \\
           Ter9  &   0.73$\pm$ 0.24  &   0.18$\pm$ 0.09& 0.92& 2.29& 0.75$\pm$0.14 &  69.96$\pm$ 2.58 \\
          Djor2   &  1.83$\pm$ 0.46  &   0.82$\pm$ 0.37& 0.17& 0.32& 0.55$\pm$0.19 &  11.51$\pm$ 6.50 \\
        NGC6517   &  2.37$\pm$ 0.37  &   0.50$\pm$ 0.15& 0.07& 0.21& 0.79$\pm$0.07 &  58.10$\pm$ 5.92 \\
        NGC6522   &  0.68$\pm$ 0.17  &   0.28$\pm$ 0.20& 0.47& 0.89& 0.59$\pm$0.25 &  71.88$\pm$123.80 \\
          Ter10   &  1.68$\pm$ 0.29 &    0.94$\pm$ 0.23& 0.26& 0.41& 0.44$\pm$0.13 &  58.94$\pm$ 0.74 \\
        NGC6535   &  2.74$\pm$ 0.16  &   1.01$\pm$ 0.23& 0.29& 0.58& 0.63$\pm$0.07 & 161.39$\pm$27.30 \\
        NGC6528   &  1.01$\pm$ 0.44  &   0.41$\pm$ 0.10& 0.21& 0.43& 0.59$\pm$0.19 & 65.60$\pm$ 1.95 \\
        NGC6539   &  2.66$\pm$ 0.13  &   1.98$\pm$ 0.16& 0.17& 0.20& 0.26$\pm$0.05 &  56.14$\pm$ 5.48 \\
        NGC6540   &  2.12$\pm$ 0.31  &   1.43$\pm$ 0.37& 0.21& 0.28& 0.33$\pm$0.14 &  22.44$\pm$ 1.48 \\
        NGC6544   &  3.05$\pm$ 0.19  &   0.62$\pm$ 0.20& 0.11& 0.33& 0.80$\pm$0.06 &  66.51$\pm$ 1.67 \\
        NGC6541   &  2.70$\pm$ 0.14  &   1.76$\pm$ 0.21& 0.10& 0.14& 0.35$\pm$0.06 &  46.22$\pm$ 4.45 \\
    ESO280-SC06   &  9.75$\pm$ 1.45  &   3.35$\pm$ 1.77& 0.28& 0.55& 0.66$\pm$0.16 &  61.52$\pm$ 3.56 \\
        NGC6553   &  1.82$\pm$ 0.15  &   1.29$\pm$ 0.22 &0.13& 0.17& 0.29$\pm$0.09 &  13.70$\pm$50.10 \\
     2MASS-GC02   &  1.46$\pm$ 0.27  &   0.48$\pm$ 0.22& 0.32& 0.77& 0.67$\pm$0.14 & 170.20$\pm$78.55 \\
        NGC6558   &  1.17$\pm$ 0.42  &   0.58$\pm$ 0.10& 0.24& 0.42& 0.50$\pm$0.19 &  62.45$\pm$22.78 \\
         IC1276   &  4.61$\pm$ 0.31  &   3.47$\pm$ 0.12& 0.17& 0.21& 0.25$\pm$0.05 &  10.88$\pm$43.39 \\
          Ter12   &  4.41$\pm$ 0.26  &   2.99$\pm$ 0.26 &0.12& 0.15& 0.32$\pm$0.05 &  18.98$\pm$ 1.82 \\
        NGC6569   &  2.39$\pm$ 0.63  &   1.84$\pm$ 0.92 &0.12& 0.14& 0.23$\pm$0.27 &  26.98$\pm$47.90 \\
          BH261   &  1.94$\pm$ 0.36  &   1.30$\pm$ 0.55 &0.13& 0.17& 0.33$\pm$0.20 &  34.28$\pm$ 0.83 \\
        NGC6584   & 10.68$\pm$ 3.70  &   2.10$\pm$ 1.08 &0.08& 0.24& 0.80$\pm$0.11 &  52.19$\pm$ 3.19 \\
        NGC6624  &   1.01$\pm$ 0.09  &   0.46$\pm$ 0.14& 0.15& 0.28& 0.54$\pm$0.11 &  68.41$\pm$ 7.54 \\
        NGC6626   &  1.74$\pm$ 0.13  &   0.57$\pm$ 0.10 &0.09& 0.20& 0.67$\pm$0.05 &  60.27$\pm$ 3.62 \\
        NGC6638   &  1.67$\pm$ 0.45  &   0.40$\pm$ 0.13& 0.11& 0.34&0.76$\pm$0.09  & 77.99$\pm$ 3.32 \\
        NGC6637  &   1.40$\pm$ 0.32  &   0.73$\pm$ 0.54 &0.16& 0.26& 0.48$\pm$0.29 &  74.26$\pm$ 1.27 \\
        NGC6642  &   1.24$\pm$ 0.16  &   0.37$\pm$ 0.15 &0.17& 0.45& 0.70$\pm$0.11 &  62.26$\pm$ 8.94 \\
        NGC6652  &   2.15$\pm$ 0.54  &   0.65$\pm$ 0.49& 0.09& 0.22& 0.70$\pm$0.20 &  74.62$\pm$ 0.91 \\
        NGC6656  &   6.20$\pm$ 0.03  &   2.96$\pm$ 0.05& 0.07& 0.11& 0.52$\pm$0.01 &  33.66$\pm$ 1.91 \\
           Pal8   &  4.01$\pm$ 0.74  &   2.44$\pm$ 1.11& 0.28& 0.38& 0.39$\pm$0.21 &  22.17$\pm$ 5.51 \\
        NGC6681   &  2.91$\pm$ 0.15  &   0.84$\pm$ 0.08& 0.10& 0.23& 0.71$\pm$0.03 &  91.13$\pm$ 3.71 \\

\hline
\end{tabular}
\end{table*}

\setcounter{table}{0}
\begin{table*}
\caption{\it continued.}
\label{tab:table1}
\begin{tabular}{@{}lcccccc}\hline

ID  & $a$      &  $R_{peri}$    & $rh/r_{J_{a}}$ & $rh/r_{J_{Rperi}}$ & $\epsilon$ & $i$ \\
     & (kpc)   &  (pc)          &                    &     &   & (deg)                  \\\hline
        NGC6712  &   2.61$\pm$ 0.07  &   0.45$\pm$ 0.10 &0.16& 0.57& 0.83$\pm$0.04 &  83.31$\pm$ 6.76 \\
        NGC6715   & 24.76$\pm$ 1.08  &  12.58$\pm$ 0.47 &0.02& 0.03& 0.49$\pm$0.03 &  83.60$\pm$ 0.55 \\
        NGC6717   &  1.81$\pm$ 0.15  &   0.89$\pm$ 0.17 &0.35& 0.60& 0.51$\pm$0.08 &  34.60$\pm$ 6.62 \\
        NGC6723   &  2.46$\pm$ 0.08  &   2.08$\pm$ 0.14 &0.16& 0.17& 0.15$\pm$0.03 &  86.46$\pm$ 9.56 \\
        NGC6749   &  3.34$\pm$ 0.19   &  1.60$\pm$ 0.28 &0.21& 0.33& 0.52$\pm$0.07 &   3.59$\pm$122.04 \\
        NGC6752   &  4.30$\pm$ 0.04  &   3.23$\pm$ 0.08 &0.08& 0.10& 0.25$\pm$0.01 &  25.96$\pm$ 2.85 \\
        NGC6760   &  3.79$\pm$ 0.11  &   1.90$\pm$ 0.09 &0.10& 0.15& 0.50$\pm$0.02 &   6.80$\pm$75.96 \\
        NGC6779   &  6.68$\pm$ 0.43  &   0.97$\pm$ 0.38 &0.10& 0.34& 0.85$\pm$0.05 & 106.58$\pm$ 8.68 \\
           Ter7   & 28.93$\pm$ 9.38  &  13.14$\pm$ 2.68 &0.13& 0.22& 0.55$\pm$0.16 &  84.81$\pm$ 0.64 \\
          Pal10   &  5.52$\pm$ 0.21  &   4.01$\pm$ 0.31 &0.10& 0.12& 0.27$\pm$0.04 &   7.62$\pm$40.80 \\
           Arp2   & 39.67$\pm$10.04  &  18.46$\pm$ 3.05 &0.15& 0.27& 0.53$\pm$0.13 &  76.93$\pm$ 1.25 \\
        NGC6809  &   3.57$\pm$ 0.08  &   1.59$\pm$ 0.02 &0.17& 0.27& 0.55$\pm$0.01 &  67.31$\pm$ 1.06 \\
           Ter8  &  35.05$\pm$10.66  &  16.23$\pm$ 3.06 &0.15& 0.27& 0.54$\pm$0.15 &  82.98$\pm$ 0.80 \\
          Pal11   &  7.30$\pm$ 1.21  &   5.43$\pm$ 2.07 &0.22& 0.27& 0.26$\pm$0.19 &  26.14$\pm$11.54 \\
        NGC6838   &  5.93$\pm$ 0.03  &   4.77$\pm$ 0.06 &0.15& 0.18& 0.19$\pm$0.01 &  11.87$\pm$11.19 \\
        NGC6864   & 10.02$\pm$ 1.50  &   2.06$\pm$ 1.15 &0.03& 0.08& 0.79$\pm$0.11 &  49.76$\pm$ 8.30 \\
        NGC6934   & 21.06$\pm$ 5.32  &   2.60$\pm$ 1.12 &0.04& 0.16& 0.88$\pm$0.06 &  23.41$\pm$46.62 \\
        NGC6981   & 12.65$\pm$ 2.29  &   1.29$\pm$ 0.74 &0.09& 0.41& 0.90$\pm$0.06 &  67.90$\pm$35.70 \\
        NGC7006  &  28.76$\pm$ 2.84  &   2.07$\pm$ 0.94 &0.06& 0.34& 0.93$\pm$0.03 & 135.42$\pm$28.56 \\
        NGC7078  &   6.98$\pm$ 0.05  &   3.57$\pm$ 0.06 &0.04& 0.06& 0.49$\pm$0.01 &  28.59$\pm$ 1.69 \\
        NGC7089   &  8.68$\pm$ 0.30  &   0.56$\pm$ 0.10 &0.04& 0.27& 0.94$\pm$0.01 &  84.10$\pm$28.49 \\
        NGC7099   &  4.82$\pm$ 0.25  &   1.49$\pm$ 0.05 &0.11& 0.23& 0.69$\pm$0.02 &118.50$\pm$ 6.29 \\
          Pal12   & 43.46$\pm$18.94  &  15.75$\pm$ 1.92 &0.14& 0.28& 0.64$\pm$0.16 &  67.35$\pm$ 3.08 \\
          Pal13   & 38.26$\pm$ 5.50 &    9.04$\pm$ 1.74 &0.26& 0.74& 0.76$\pm$0.05 & 112.26$\pm$ 6.36 \\
        NGC7492  &  16.25$\pm$ 1.88 &    4.27$\pm$ 2.29 &0.18 &0.45& 0.74$\pm$0.12 &  91.70$\pm$ 1.20 \\
\hline
\end{tabular}
\end{table*}


\bsp	
\label{lastpage}
\end{document}